\documentclass{ws-ijmpa}
\usepackage{graphicx}
\usepackage{epsfig}
\usepackage{amsfonts}

\usepackage{amsmath, amssymb, array}
\usepackage{graphics,graphicx,float,mfpic}
\usepackage{epsfig}
\usepackage{color}
\newcommand{\be}{\begin{equation}}
\newcommand{\ee}{\end{equation}}
\newcommand{\bea}{\begin{eqnarray}}
\newcommand{\eea}{\end{eqnarray}}
\newcommand{\bmat}{\begin{pmatrix}}
\newcommand{\emat}{\end{pmatrix}}

\begin{document}

\markboth{T\'erence Delsate}
{Stability of AdS black strings}

%
\catchline{}{}{}{}{}
%

\title{STABILITY OF ADS BLACK STRINGS}

\author{T\'erence Delsate}

\address{Theoretical and Mathematical Physics Department, \\Universit\'e de Mons-Hainaut, 20, Place du Parc\\
7000 Mons,
Belgium\\
terence.delsate@umh.ac.be}

\maketitle

\begin{history}
\received{Day Month Year}
\revised{Day Month Year}
\end{history}

\begin{abstract}
We review the recent developements in the stability problem and phase diagram for asymptotically locally $AdS$ black strings. 
First, we quickly review the case of locally flat black string before turning to the case of locally $AdS$ spacetimes.

\end{abstract}

\ccode{PACS numbers: 04.20.-q, 04.50.Gh, 04.70.Bw}

\section{Introduction}
This decade has witnessed a growing interest for solutions of general relativity in $AdS$ spaces. This is due to the celebrated $AdS/CFT$ correspondance conjecture\cite{adscft}, relating solutions of general relativity in asymptotically $AdS$ spaces to conformal field theories defined on the conformal boundary of the $AdS$ space. In this context, black hole solutions play an important role\cite{adscftbh}. 

In more than four dimensional spacetime, the uniqueness theorem on black holes, garantying that the horizon topology of a black object is always $S^2$ is no longer true. Various black objects have been constructed in higher dimensions, such as black strings with horizon topology $S^{d-3}\times S^1$ in contrast with black holes  with horizon topology $S^{d-2}$.

On the other hand, in 1993, R. Gregory and R. Laflamme have shown that black strings and branes are unstable towards long wavelength perturbations\cite{gl}. The Gregory-Laflamme instability was originally discovered in the framework of asymptotically locally flat spacetimes but it is believed to be a generic feature of black extended objects. In particular, it will be argued that this instability persists in asymptotically locally $AdS$ spacetimes, where a black string solution has been found recently by R. Mann, E. Radu and C. Stelea\cite{rms}.

This proceeding is organised as follows: we review the black string instability and phase diagram in asymptotically locally flat spacetimes in section 2 before turning to asumptotically locally $AdS$ spacetimes in section 3.

\section{Asymptotically locally flat space}
Thoughout this section, we consider the $d$-dimensional Einstein-Hilbert action
\be
S = \frac{1}{16\pi G}\int_{\mathcal M} \sqrt{-g} R d^dx+ \frac{1}{8\pi G}\int_{\partial\mathcal M} \sqrt{-h}K d^{d-1}x,
\ee
where $G$ is the $d$-dimensional Newton constant which we set to one, $\mathcal M$ is the spacetime manifold, $g$ is the determinent of the metric, $R$ is the scalar curvature and $K$ is the extrinsic curvature of the boundary manifold $\partial\mathcal M$. The equation of motion resulting form the variation of the Einstein-Hilbert action is given by
\be
R_{MN} = 0,\ M,N=0,\ldots,d-1,
\label{eqflat}
\ee
where $R_{MN}$ is the Ricci tensor.

The black string solution to equation \eqref{eqflat} is given by
\be
ds^2 = -\left( 1 - \left(\frac{r_0}{r}\right)^{d-4} \right)dt^2 + \frac{dr^2}{\left( 1 - \left(\frac{r_0}{r}\right)^{d-4} \right)} + r^2d\Omega_{d-3}^2 + dz^2,\ z\in[0,L],
\label{flatbs}
\ee
where $L$ is the length of the coordinate $z$, $r_0$ is the horizon radius and $d\Omega_{d-3}$ is the line element of the unit $(d-3)$-sphere.

The black string \eqref{flatbs} is characterised by thermodynamical quantities, namely the mass $M$, associated to the time translation, the tension $\mathcal T$, associated with $z$-translation invariance, the temperature $T_H$, which can be obtained by demanding regularity at the horizon in the euclidean section, the entropy $S$, defined as one quarter of the horizon area and the length in the extradimension $L$.

These thermodynamical quantities can be used to define a thermodynamical phase diagram in temperature-entropy coordinates $(T_H,S/L)$ or in a mass-tension diagram $(\mu,n)$, with the dimensionless quantities $\mu= M/L^{d-3}, n= \mathcal TL/M$\cite{ho}. 

The uniform black string solution is subject to a dynamical instability which manifests itself already at the linearised level\cite{gl}. The equations for the perturbations admit unstable solutions with small wavenumber $k$ in the extradirection as well as stable solutions with a large wavenumber. The wavelengths of the various modes are given by $\lambda = 2\pi/k$. There is a static solution between these two regimes for $k=k_c$, where $k_c$ is called the critical wavenumber. 

This dynamical instability is related to the thermodynamical stability of the black string: long black strings are unstable while short black strings are stable
We refer the reader to the original paper\cite{gl} for more details.

Initially, it was widely believed that the unstable black string should decay to an array of localised black holes but it has been shown that this decay would take an infinite proper time at the horizon\cite{horowitz}. This suggested the existance of another phase, the non uniform black string. Non uniform black strings were first constructed in a perturbative way then by solving numerically the full system of non-linear partial differential equations\cite{gubser}. All these three phases, the black string, non uniform black string and the localised black hole are static solution; these static solutions should be the equilibrium configurations since they don't evolve by definition.

A possible way to have an idea of the endpoint of the black string instability consists in comparing the thermodynamical properties of the three phases in a phase diagram\cite{ho}.

\section{Asymptotically $AdS$ space}
In this section, we present the recent results obtained in the stability problem for black strings in $AdS$ spacetime. We consider the $d$-dimensional Einstein-Hilbet action with a negative cosmological constant,
\be
S = \frac{1}{16\pi G}\int_{\mathcal M} \sqrt{-g} \left(R+\frac{(d-1)(d-2)}{\ell^2}\right) d^dx+ \frac{1}{8\pi G}\int_{\partial\mathcal M} \sqrt{-h}K d^{d-1}x,
\label{actads}
\ee
with the same convention as in the previous section and where $\ell$ is the $AdS$ radius, related to the cosmological constant $\Lambda$ by $\Lambda=-(d-1)(d-2)/2\ell^2$.

The uniform black string solution is obtained by using the spherically symmetric ansatz
\be
ds^2 = -b(r)dt^2 + \frac{dr^2}{f(r)} + r^2d\Omega_{d-3}^2 + a(r)dz^2
\label{adsbs}
\ee
and solving numerically the equations of motion resulting from the variation of \eqref{actads}\cite{rms}. The asymptotic behaviour of the metric fields is given by $a,b,f\approx r^2/\ell^2$ at the leading order.

The thermodynamical quantities characterising the solution \eqref{adsbs} are the same as in the asymptotically locally flat case, but have to be computed in a regularised version of the action which diverges because of the $AdS$ asymptotic. The regularised action is obtained by adding appropriate boundary counterterms\cite{bala}.

However, there is a new lengthscale in the theory, namely the $AdS$ radius which affects the thermodynamical properties of the black string. In particular, it allows a new phase of thermodynamically stable black string, namely big $AdS$ black strings, characterised by a large horizon radius $r_0$ - $AdS$ length ratio. Small $AdS$ black strings, with $r_0/\ell<<1$, are thermodynamically unstable.

We investigated the existance of a Gregory-Laflamme instability by considering non uniform black strings within the ansatz
\be
ds^2 = -b(r)e^{2A(r,z)} + e^{2B(r,z)}\left(\frac{dr^2}{f(r)} + a(r)dz^2\right) + e^{2C(r,z)}d\Omega_{d-3}^2,\ z\in[0,L],
\ee
where $a,b,f$ is the solution of Mann, Radu and Stelea and $A,B,C$ are smooth functions of $r$ and $z$. In the perturbative approach, we develop the non uniformity in a Fourier series of the variable $z$ and in term of a small parameter $\epsilon$ according to $X(r,z)=\epsilon X_1(r)\cos(kz)+ \epsilon^2(X_0(r) + X_2(r)\cos(2kz)) + \mathcal O(\epsilon)^2 $, $X$ generically denoting $A,B,C$. Then we solve at each order in $\epsilon$ the corresponding equations of motion. Order $\epsilon^0$ is just the uniform solution while order $\epsilon$ gives access to the linear stability problem. The modes $X_1$ correspond to the static perturbation in the Gregory-Laflamme picture. The equations for these modes form an eigenvalue problem, where the eigenvalue is the square of the Gregory-Laflamme critical wavenumber. If the eigenvalue is real, there exists a Gregory-Laflamme instability, if it is imaginary, the solution is always stable\cite{rbd}. It turns out that small $AdS$ thermodynamically unstable black strings are dynamically unstable while big $AdS$ thermodynamically stable black strings are dynamically stable, confirming the Gubser-Mitra conjecture\cite{gm} in this case.

The order $\epsilon^2$ contains two independant modes: the backreactions, $X_0$ and massive modes $X_2$. Thermodynamical corrections for the non uniform phase arise at this order where only the backreactions contribute\cite{pnubsads}. We investigated the non uniform solutions emanating from the unstable modes at the order $\epsilon^2$ for small $AdS$ black strings. The new lengthscale provided by $\ell$ implies a new dimensionless quantity, $\mu_2=L/\ell$ characterising the non uniform solutions. It turns out that for small value of $\mu_2$, the picture is essentially similiar to the case of asymptotically locally flat spacetimes while for large $\mu_2$ enough, the small $AdS$ phase becomes thermodynamically stable. In other words, the length $L$ plays an important role in the thermodynamical stability of non uniform black strings, just like $r_0$ does for uniform black strings; small ($r_0/\ell<<1$) and short ($L/\ell<<1$) $AdS$ non uniform black string are thus thermodynamically unstable while small and long ($L/\ell \approx 1$) non uniform black strings are thermodynamically stable\cite{pnubsads}.

\section{Acknowledgement}
I would like to thank the organisers of the 7th Friedman Seminar for giving me the opportunity of presenting this talk. I also would like to thank Yves Brihaye, Eugen Radu and Georges Kohnen for usefull discussions and advices.

\end{document}